\documentclass[twocolumn]{aastex701} 

\usepackage{multirow}

\usepackage{graphics,epsf}
\usepackage[utf8]{inputenc}
\usepackage{amsmath}                
\usepackage{amsfonts}               
\usepackage{amssymb}                
\usepackage{epsfig}                 
\usepackage{graphicx}               
\usepackage{float}
\usepackage{color}
\usepackage{multirow}               

\hypersetup{
    colorlinks=true,
    linkcolor=red,   
    urlcolor=cyan}

\usepackage[colorinlistoftodos]{todonotes}

\newcommand{\cm}{{~\rm cm}}
\newcommand{\km}{{~\rm km}}
\newcommand{\s}{{~\rm s}}

\newcommand{\g}{{~\rm g}}
\newcommand{\G}{{~\rm G}}
\newcommand{\K}{{~\rm K}}

\newcommand{\yr}{{~\rm yr}}

\newcommand{\mum}{{~\rm \mu m}}

\begin{document}
\title{Supernova 1987A was a ``failed supernova'' twenty thousand years before its jet-driven explosion}

\author[0000-0003-0375-8987]{Noam Soker}
\affiliation{Department of Physics, Technion - Israel Institute of Technology, Haifa, 3200003, Israel; soker@technion.ac.il}
\email{soker@physics.technion.ac.il}

\begin{abstract}
I show that the progenitor of supernova (SN) 1987A faded substantially and became red for hundreds of years to observers in and near the plane of the equatorial ring, because the equatorial ring, which was formed by a binary interaction between the SN 1987A progenitor and a main-sequence companion that ejected material about 20,000 years before the explosion, obscured the progenitor. For a few hundred years, the luminosity for such hypothetical observers was mainly light scattered by the two outer rings, which amounted to $\simeq 5\%$ of the progenitor luminosity. In present terms, this event would have been classified as a ``failed supernova’’ by these observers, although there was no core collapse, no black hole formation, and nothing failed. Rather, this event was a type II ILOT (intermediate-luminosity optical transient), in which equatorial ejecta temporarily obscured the central source. After a few hundred years, the outer rings became transparent, the inner ring became more transparent, and the SN 1987A progenitor brightened over a few thousand years to its normal luminosity for equatorial observers. This finding strengthens the claim that ``failed supernova’’ candidates are likely to be type II ILOTs.  Although the binary interaction also spun up the core, the bipolar morphology of SN 1987A (its Keyhole) is misaligned with the triple-ring system. I use the jittering jets explosion mechanism (JJEM) to speculate on a scenario that might explain this misalignment. This study adds small but unique support to the claim that the JJEM is the primary explosion mechanism of CCSNe.
\end{abstract}

\keywords{stars: massive -- supernovae: general -- (stars:) supernovae: individual (1987A) -- stars: jets -- ISM: supernova remnants}  

\section{Introduction} 
\label{sec:intro}

A fierce and aggressive debate exists over the explosion mechanism of core-collapse supernovae (CCSNe), including the iconic CCSN remnant (CCSNR) SN 1987A. Some researchers claim the jittering-jet explosion mechanism (JJEM), while others support the neutrino-driven (neutrino-heating) mechanism. The observables during the explosion of SN 1987A, such as progenitor properties, neutrino emission, explosion energy, spectra, and the light curve, cannot distinguish between the two theoretical explosion mechanisms (e.g., \citealt{Soker2025Learning}). 
Therefore, studies compare SN 1987A's morphology with their preferred theoretical predictions, whether the neutrino-driven mechanism (e.g., \citealt{Jankaetal2017, Alpetal2019, Jerkstrandetal2020, Galberetal2021, Wessonetal2026}), or the JJEM (e.g., \citealt{Soker2017RAATwo, Soker2024NA1987A, Soker2024PNSN, Soker2026Dust, Soker2026SN1987Amulecular, BearSoker2018}). 
Some simulations also highlight the bipolar geometry of SN 1987A (e.g., \citealt{Onoetal2020, Orlandoetal2020, Orlandoetal2025}). 
For recent reviews of the JJEM see \cite{Soker2024UnivReview, Soker2025Learning}, and for recent papers \cite{ShiranSoker2025, Soker2026J0450, WangShishkinSoker2025, Soker2026Long}. For recent reviews of the neutrino-driven mechanism (the delayed neutrino mechanism), see, e.g., \cite{Janka2025, Janka2025Padova}, and for recent papers, see, e.g., \cite{Calvertetal2026, Giudicietal2026, Karimetal2026, LuoZhaKajino2026, Mezzacappa2026, Murphyetal2026, Orlando2026, Rusakovetal2026, VarmaMuller2026}. The magnetorotational explosion mechanism, which explains CCSNe with one pair of jets, applies only to rare types of CCSNe and does not attempt to explain most CCSNe (for a recent paper on this mechanism, see, e.g., \citealt{Celatietal2026}). 
        
Another dispute between the supporters of the neutrino-driven mechanism and the JJEM is whether ``failed supernovae'' exist. 
The neutrino-driven mechanism predicts that a non-negligible fraction of massive stars collapse to a black hole without explosion, an event resulting in a faint transient (e.g., \citealt{AntoniQuataert2023}), and term this a ``failed supernova;'' some black holes do form in a CCSN according to the neutrino-driven mechanism (e.g., \citealt{BurrowsWangVartanyan2025}).   
In contrast, according to the JJEM, no failed CCSNe exist, and all massive stars experience jet-driven explosions, even when forming black holes. The dispute centers on a few specific observed cases of fading massive stars.  
   
For the fading event N6946-BH1 (\citealt{Gerkeetal2015}; \citealt{Humphreys2019} argued it was a yellow hypergiant), \cite{Adamsetal2017} (also \citealt{Kochanek2024, Kochaneketal2024}) claimed for a ``failed supernova''. At the same time, we 
\citep{KashiSoker2017, SokerTypeII2021, BearetalTypeII2022} proposed a type II ILOT (intermediate luminosity optical transient) event. In a type II ILOT, strong binary interaction ejects dusty material in a small solid angle towards the observer; this dusty ejecta causes the temporary fading. \cite{Beasoretal2024} argued that the type II ILOT scenario is compatible with their detection of a luminous infrared emission from N6946-BH1.
    
\cite{Deetal2026SciA} and \cite{Deetal2026B} (also \citealt{Antonietal2025}) proposed that  M31-2014-DS1 was a ``failed supernovae.''  I \citep{Soker2024UnivReview, Soker2026FailedFailed}, on the other hand, analyzed this event in detail and concluded that it fits a type II ILOT much better. \cite{Beasoretal2026} studied this event observationally and found the type II ILOT scenario to fit its properties better.  \cite{Beasoretal2025} claimed a non-spherical dust distribution, as predicted by the type II ILOT scenario.  

Studies of large populations of red supergiants concluded that only a small number, if any, of failed supernovae exist  \citep{ByrneFraser2022, StrotjohannOfekGalYam2024, Beasoretal2025, Healyetal2025}.  \cite{Beasoretal2025} argued that the inferred luminosity of many CCSN progenitors is underestimated, implying their masses are $ > 20 M_\odot$; this reduces the need to assume failed CCSN, and makes the number of failed CCSNe very low, or even zero.  

On the other hand, evolved massive stars can lose mass at high rates and in complicated morphologies (e.g., \citealt{Michaelisetal2018, Gilkisetal2025, Grichener2025,  MukhijaKashi2025ApJ, MukhijaKashi2026ApJa, MukhijaKashi2026ApJb, MukhijaKashi2026NewA, MukhijaKashi2026PASP}). Such mass loss episodes, if they cover a small fraction of the sky and are towards the observer, can lead to a type II ILOT. 
  
In this study, I show that the equatorial ring the progenitor of SN 1987A ejected $\simeq 20,000 \yr$ before its explosion has obscured SN 1987A from any hypothetical observer along its expansion direction, leading to the appearance of a ``failed supernova'' (Section \ref{sec:Failed}). I then discuss the misalignment between the axis of the three circumstellar rings and the axis of the explosion's bipolar ejecta (Section \ref{sec:Misalignedaxes}). I account for this misalignment within the JJEM framework. In Section \ref{sec:Summary}, I summarize this study, another in a series of papers on SN 1987A in the JJEM framework.     

\section{The rings and a ``failed supernova''} 
\label{sec:Failed}

\subsection{Only an equatorial ring} 
\label{subsec:OnlyRing}

\begin{figure} [t]
\centering
\includegraphics[trim=0.5cm 15.8cm 0.0cm 4.3cm ,clip, angle=0, scale=0.55]{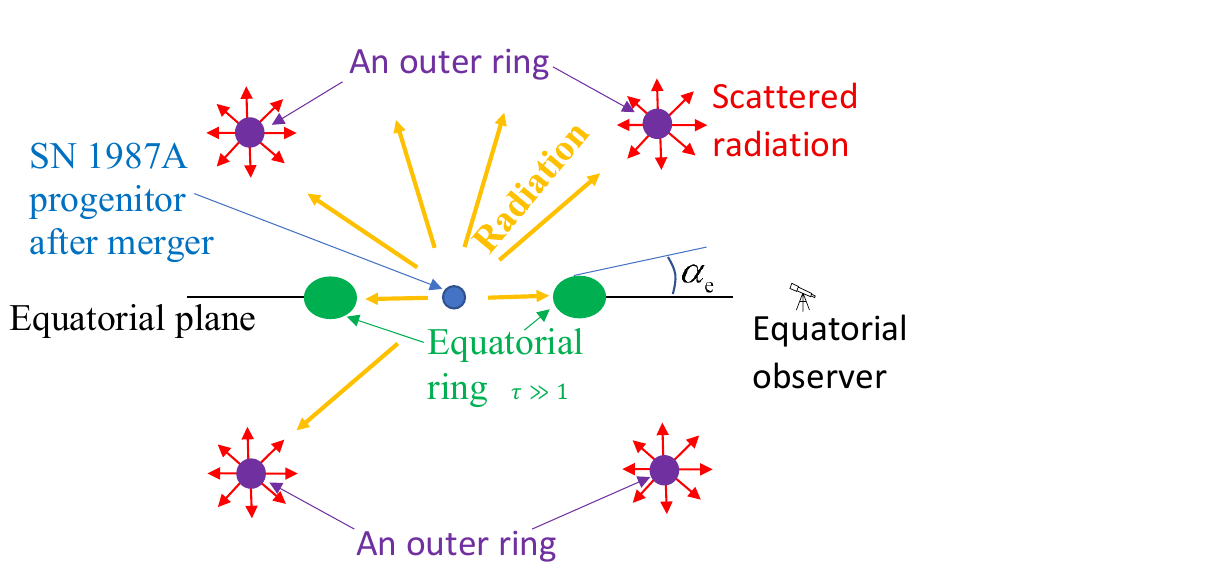}
\caption{A schematic drawing in the meridional plane (perpendicular to the equatorial plane and through the center) of the three rings of SN 1987A that lead to a type II ILOT at about ten to hundreds of years after the merger (the common envelope evolution). A hypothetical observer in the plane of the equatorial ring receives mainly radiation scattered by the outer rings, amounting to $\approx 5 \%$ of the central star luminosity. For another few thousand years, the equatorial ring becomes more transparent, and the luminosity rises back to the central star luminosity.   
}
\label{Fig:figure1}
\end{figure}

In \cite{SokerTypeII2021}, I considered a type II ILOT that ejects a mass $M_{\rm ring,e}$ with a velocity $v_{\rm e}$ into an expanding equatorial ring within a half-opening angle $\alpha_e$. The ejection process lasts for a time of $\Delta t_{\rm e}$, from $t=-\Delta t_{\rm e}$ to $t=0$. 
I derived a very crude relation between the luminosity as inferred by an equatorial observer (i.e., one within an angle $\alpha_e$ from the equatorial plane), $L_{\rm e}$, and the total ILOT luminosity, as a function of time.  
\begin{eqnarray}
\begin{aligned}
& \frac{L_{\rm e}} {L_{\rm I}}
 \approx 0.003
\left( \frac{\sin \alpha_e}{\sin 15^\circ} \right)^{5/2}
\left( \frac{\tau_{{\rm T},z}}{\tau_{{\rm T},r}} \right)^{3/2}
\left( \frac {M_{\rm ring,e}}{0.1 M_\odot} \right)^{-1/2}
\\ & \times
\left( \frac {v_{\rm e}}{10 \km \s^{-1}} \right)
\left(\frac{t+ \Delta t_{\rm e}}{33 \yr}\right)^{1/2}
\left(\frac{t}{30 \yr}\right)^{1/2}. 
\label{eq:Ltot}
\end{aligned}
\end{eqnarray}
Here $\tau_{{\rm T},z}$ and $\tau_{{\rm T},r}$ are the Thomson optical depth in the disk perpendicular to the orbital plane and along the orbital plane (radially), respectively. For a torus, as is assumed here for the equatorial ring of SN 1987A, $\tau_{{\rm T},z} \simeq \tau_{{\rm T},r}$. In Figure \ref{Fig:figure1}, I present a schematic drawing of the rings in a meridional plane through the SN 1987A progenitor, at about ten to a few hundred years after the merger.

The equatorial ring of SN 1987A expands at a (pre-explosion) velocity of $v_{\rm e} = 10.3 \km \s^{-1}$ \citep{CrottsHeathcote1991}, and its mass is  $M_{\rm ring,e} \simeq 0.058 M_\odot$ (e.g., \citealt{Mattilaetal2010}).

The radius of the ring (from the explosion to the ring) is $0.858 \pm 0.011^{\prime \prime}$ and its width is $0.121 \pm 0.022^{\prime \prime}$, or an inner radius of $\simeq 60.5^{\prime \prime}$ (e.g., \citealt{Plaitetal1995}). From these, $\sin \alpha_e = 60.5/\sqrt{858^2+60.5^2}=0.07$, implying a half opening angle of $4^\circ$. However, the intensity in the ring decreases from its radius to the two sides and does not end abruptly (e.g., \citealt{Plaitetal1995}). I therefore use the scaling with $\alpha_e = 15^\circ$. 

The total luminosity in the waveband $\lambda<5\mum$ is less than 10\% of the total luminosity inferred in the equatorial plane, $L_{\rm e}$, as given by equation (\ref{eq:Ltot}). 

\subsection{The outer rings} 
\label{subsec:OuterRings}

The three rings are the densest parts of a bipolar structure (e.g., \citealt{Sugermanetal2005}). At early times after the ejection of the rings and the bipolar structure, a large fraction of the radiation escaped along the polar cones. However, this is not the morphology of an equatorial ring. The bipolar ejecta was bright in the IR, but only a fraction $\approx 0.5$ of the central source luminosity was visible to an observer in the equatorial plane. Later, depending on the density and mass between the rings, this material becomes transparent as the wind from the central star accelerates it outward, forming the structure of the three rings.   

From the images of the outer rings and their analysis (e.g., \citealt{Panagiaetal1996, Sugermanetal2005, Tziamtzisetal2011}), their average angle from the symmetry axis through the rings is $\theta_{\rm O} \simeq 50^\circ$.  The average radius of the cross section of the outer rings is $b_{\rm O} \simeq 0.04 a_{\rm O}$,
where $a_{\rm O}$ is the average radius of the ring. Each extends over an angle of $\Delta \theta_{\rm O} \simeq 3.5^\circ$.
The two rings together cover a solid angle of $\Omega_{\rm 2O}$, or a fraction of the sphere given by 
\begin{equation}
\frac{\Omega_{\rm 2O}}{4 \pi} \simeq \frac{2}{4 \pi}(2 \pi \sin \theta_{\rm O} \Delta \theta_{\rm O}) \simeq 0.05. 
    \label{eq:OmegaOuter}
\end{equation}
As long as they are optically thick, the outer ring absorbs and scatters the central-source radiation and emits it toward the observer (mainly in the IR). This implies that the bolometric luminosity to a hypothetical observer in the equatorial plane of the inner ring of SN 1987A would be $\simeq 5 \%$ of the central source luminosity (as the contribution from the equatorial ring is much smaller by equation \ref{eq:Ltot}).     

The typical total number density in the outer rings is $\simeq 10^3 \cm^{-3}$ (e.g., \citealt{Rosuetal2026}), and their distance from the center is $R_{\rm O} \simeq 1.5 \times 10^{18} \cm$. The width of the rings is $\simeq 10^{17} \cm$, and the mass of each ring is $M_{\rm O} \simeq 0.02 M_\odot$. The optical depth of the outer rings to Thomson scattering is $\tau_{\rm O,T} \simeq 5 \times 10^{-5}$. 
The optical depth in the case of dust can be an order of magnitude larger: $\tau_{\rm O,d} \approx  10^{-3}$. The outer rings became optically thin, $\tau \simeq 1$, when they were $(\tau_{\rm O})^{-1/2}$ times closer to the center, or, with an age of $2\times 10^{4} \yr$, at about $\simeq 100-500 \yr$ after the ejection of the triple ring system. The inner ring, which is $\approx 100$ times denser, and also wider, became optically thin only much later. The material between the rings might have become optically thin within several years after the end of the ejection of the ring system.  

\subsection{The ``failed supernova'' event} 
\label{subsec:FailedEvent}

From the previous subsections I estimate the following light curve of SN 1987A to a hypothetical observer in the direction of the equatorial ring in the years following the ejection of the triple ring bipolar structure. 

Shortly after the ejection, the ejecta was optically thick in most directions, besides the polar cones, and the source was fainter than the central source, but by a factor of a few. Within a time of $\approx 10 \yr$, the material between the three rings became optically thin. From that time to about hundreds of years until the outer rings have turned optically thin, an observer near the equatorial plane, along the direction of the inner ring, has received a very small amount of radiation from the equatorial ring (equation \ref{eq:Ltot}). It would have received mainly radiation processed by the two outer rings. Since they cover a fraction of $\simeq 0.05$ of the sphere's solid angle, this would be the fraction of the luminosity of the central source that the hypothetical equatorial observer would have received. This is very similar to the two ``failed supernova'' candidates, which I argue are type II ILOTs: N6946-BH1 luminosity is $13\% - 25 \%$ of its progenitor (e.g., \citealt{Beasoretal2024}), and M31-2014-DS1 luminosity is $\simeq 7 \%$ of its progenitor (e.g., \citealt{Beasoretal2026}).  

After $\simeq 100- 500 \, \yr$, the outer rings became optically thin, and their contribution to the luminosity of the SN 1987A progenitor to a hypothetical equatorial observer decreased further. However, at that time, $t \approx 500-1000 \yr$, the luminosity from the equatorial rings reached a value of $\simeq 5 \%$ (equation \ref{eq:Ltot}). From then on, the luminosity to the equatorial observer has slowly increased. A few thousand years after the ejection of the triple-ring system, the equatorial ring started to become optically thin, and luminosity increased back toward the progenitor one.  

Overall, for a hypothetical equatorial observer 20,000 years ago, the formation of the triple-ring system in SN 1987A was a Type II ILOT event, which could observationally have been interpreted as a ``failed supernova''. Of course, there was no core-collapse, no black hole formation, and nothing failed.

\section{Misaligned axes} 
\label{sec:Misalignedaxes}

\subsection{Motivation} 
\label{subsec:Motivation}

The merger of the SN 1987A progenitor with a lower-mass main-sequence star triggered the ejection of the triple-ring systems (e.g., \citealt{Soker1999, MorrisPodsiadlowski2007, MorrisPodsiadlowski2009}). The suggestion for a common envelope evolution of the SN 1987A progenitor with a main-sequence star, and then the destruction of the latter near the core of the progenitor, came from the asymmetrical explosion \citep{ChevalierSoker1989} and the blue progenitor \citep{Podsiadlowskietal1990A}. This companion has spun up the envelope of SN 1987A's progenitor. However, the asymmetry axis of the bipolar ejecta of SN 1987A and the symmetry axis of the triple-ring structure are misaligned. Below, I speculate on a scenario that might account for this misalignment within the JJEM framework. 

\cite{BearSoker2018} identified a southern jet-like molecular structure in images of SN 1987A that \cite{Abellanetal2017} presented. In \cite{Soker2026SN1987Amulecular}, I used images from \cite{Wessonetal2026} and identified the full bipolar morphology of the ejecta. I attribute it to a pair of jets (or several pairs along the same axis) in the framework of the JJEM. The axis of the bipolar ejecta is inclined to the axis of the triple-ring system by $\alpha_i \simeq 45^\circ$. If the companion that spun up the envelope also spun up the core, the naive expectation would be that the two axes are aligned. This is not the case. 

\subsection{The spun-up core} 
\label{subsec:SpunUpCore}
Consider the tidal destruction of the companion. This takes place at the radius where the average density of the SN 1987A progenitor becomes larger than the average density of the companion. Models of a zero-age main-sequence star of $20 M_\odot$ show that at $\simeq 20,000 \yr$ before core collapse, the mass in the core and very inner envelope is $M_{\rm core} \simeq 8 M_\odot$. A main-sequence companion of mass $\simeq {\rm few} \times M_\odot$ will be tidally destroyed at $R_{\rm T} \simeq 3 R_\odot$. The (baryonic) mass coordinate to be accreted last onto the neutron star, of a final gravitational mass of $M_{\rm NS}=1.4 M_\odot$, at $\simeq 20,000 \yr$ before core collapse is the radius of the mass coordinate $m\simeq 1.5 M_\odot$, $R(1.5) \simeq 0.1 R_\odot$; it would be at only a few thousand kilometers at the time of core collapse. I take the companion to spin up this mass shell to a fraction $\eta_s$ of the angular velocity at the tidal destruction radius 
\begin{equation}
 \omega(1.5) = \eta_s \left(\frac{GM_{\rm core}}{R^3_{\rm T}}\right)^{1/2}.  
    \label{eq:omegaRot}
\end{equation}
There are two reasons for $\eta_s < 1$. First, the companion will not bring the material at $R_{\rm T}$ to the breakup rotation at that radius. Second, the regions between $R(1.5)$ and $R_{\rm T}$ are radiative, so it is not clear that the region reaches a solid-body rotation. The specific angular momentum of the $m=1.5M_\odot$ shell at its equator is 
\begin{equation}
 \begin{split}
& j_{\rm e}(1.5)  = \omega(1.5) R^2 (1.5) = 
 5 \times 10^{15}  
\left(\frac{\eta_s}{0.3}\right) 
\\ & \times 
\left(\frac{M_{\rm core}}{8M_\odot} \right)^{1/2} 
\left(\frac{R_{\rm T}}{3R_\odot} \right)^{-3/2} 
\left(\frac{R(1.5)}{0.1R_\odot} \right)^{2} 
\cm^2 \s^{-1}.  
    \label{eq:je}
 \end{split}
\end{equation}

The value of the equatorial specific angular momentum is an order-of-magnitude estimate and should be considered as such. Nonetheless, it is meaningful in relation to two values of specific angular momentum.  The specific angular momentum to form a thin accretion disk around the newly born neutron star of mass $1.4 M_\odot$ and radius of $12 \km$ is $j_{\rm disk} \simeq 1.5 \times 10^{16} \cm^2 \s^{-1}$. The newly born neutron star has a larger radius, $\simeq 20-30 \km$, and the required specific angular momentum to form a disk is $\simeq 2 \times 10^{16} \cm^2 \S^{-1}$. However, as I discuss later, I also allow for an accretion belt formation; hence, I scale with $j_{\rm disk} \simeq 1.5 \times 10^{16} \cm^2 \s^{-1}$. 
In other words, the spun-up core is not sufficient to form a thin accretion disk, but might form an accretion belt (see Section \ref{subsec:LateTimes}). 
The other is the specific angular momentum parameter that previous works used (e.g., \citealt{WangShishkinSoker2024}) to characterize the specific angular momentum fluctuations in the pre-collapse convective layer in the inner core
\begin{equation}
    j_{\rm conv}=v_{\rm conv}r \approx 2.5 \times 10^{15} - 10^{16} \cm^2 \s^{-1},
    \label{eq:jParameter}
\end{equation}
where $v_{\rm conv}$ is the convection velocity and the typical values of the convective layers of the inner core ($m \lesssim 2.5 M_\odot$) are from \cite{WangShishkinSoker2025}. 

 The two crude estimates applicable to SN 1987A are 
\begin{equation}
\frac{j_e(1.5)}{j_{\rm disk}} \approx 0.3, 
\qquad {\rm and} \qquad
\frac{j_e(1.5)}{j_{\rm conv}} \approx  1 .
    \label{eq:ratios}
\end{equation}
The first ratio implies that the companion to SN 1987A's progenitor could have spun up the core of SN 1987A to a non-negligible velocity, but not one that would lead to a long-lived accretion disk along the same axis as that of the orbital angular momentum. The second ratio implies that the specific angular momentum contributions from convective fluctuations and the induced core rotation could have been of the same order of magnitude. That the rotation specific angular momentum is not much larger than the fluctuations explains why the axis would not necessarily be along that of the rings. The naive expectation would be for jittering jets. This ratio does not explain why a large fraction of SN 1987A ejecta is in a bipolar structure. 
 I turn to propose a plausible explanation for that. 

\subsection{The late directional rotational accretion} 
\label{subsec:LateTimes}

I speculate on a possible scenario to account for the bipolar ejecta of SN 1987A, its 'Keyhole,' which I attributed in previous studies to a pair of jets (e.g., \citealt{BearSoker2018, Soker2024PNSN, Soker2026Dust, Soker2026SN1987Amulecular}). 
Particularly, in \cite{Soker2024Keyhole}, I considered a long-lived jet-pair
that, during the final explosion phase, have shaped the Keyhole. In that paper, I considered a long-lived jet pair as a result of random fluctuations in angular momentum. Such fluctuations can create positive feedback: later accretion episodes are more likely to have fluctuating angular momentum with a positive component along the original angular momentum of the accretion disk that launched the jet pair than a fully random set of fluctuations. This positive feedback comes from two inclined consecutive jet-launching episodes forcing the next episode to have its angular momentum axis in the same plane as the first two, in what is called a planar jittering-jets pattern \citep{PapishSoker2014Plan}. Although that effect exists, it is unclear whether it is sufficient for SN 1987A. I therefore consider a new mechanism to launch a long-lived pair of jets at the end of the explosion process, based on pre-collapse core rotation.  

Equation (\ref{eq:ratios}) shows that the core rotation of SN 1987A's progenitor was non-negligible, although fluctuations in the angular momentum could have dominated the angular momentum. I speculate on the following plausible scenario in the JJEM framework with a moderate pre-collapse core rotation.  I term it `late directional rotational accretion.' 

 I consider the following phases of the late directional rotational accretion scenario. 
 \begin{enumerate}
     \item The early explosion process proceeds by several pairs of jittering jets. The random component of the angular momentum dominates, and there is no substantially preferred direction; there might be some preference for the initial axis of the pre-collapse core rotation, $\vec{J}_{\rm core}$. 
     \item As the explosion in the JJEM proceeds, jets explode along more and more directions through the core until the entire core is either accreted onto the newly born neutron star or explodes. Because of the point symmetry (as the neutron star launches pairs of opposite jets), in many cases the last two directions that were not exploded are on opposite sides of the center. Consider two opposite directions along a line that is inclined at an angle $\beta$ to $\vec{J}_{\rm core}$.     
     \item The material from these two directions is accreted onto the neutron star, and might have a net angular momentum that might boost the formation of a long-lived accretion disk. It originates at larger radii, as accreting late in the collapse, and might have sufficient angular momentum to build a thick accretion belt or disk.      
     \item Such a thick accretion belt or disk launched the final energetic pair of jets that shaped the Keyhole of SN 1987A. In an accretion belt, the specific angular momentum of the accreted gas is below the one required to orbit the central object. However, it is still large enough to leave empty funnels along the two polar directions. Such an accretion belt might launch jets in CCSNe (e.g., \citealt{SchreierSoker2016}).      
 \end{enumerate}

I elaborate on phase 3 of the late directional rotational accretion. 
Consider that the two opposite accreted sides are at an angle $\beta$ to the pre-collapse rotation. They rotate around the axis with a velocity of $v_{\rm dir} = \omega (r \sin \beta)$. The two accreted sides have opposite velocity directions and opposite sides, hence the same sense of angular momentum. Their velocity is perpendicular to the radius, and hence the specific angular momentum is 
\begin{equation}
j_{\rm dir}=v_{\rm dir} r = \omega r^2 \sin \beta \simeq j_{\rm e} (1.5) \sin \beta. 
    \label{eq:jdir}
\end{equation}
The angular momentum direction of the two accreted directions is not along the pre-collapse core angular momentum, but rather at an angle of $90^\circ - \beta$ to it.  
To account for the Keyhole, $90^\circ - \beta = \alpha_i \simeq 45^\circ$, i.e., $\beta \simeq 45^\circ$.

Late directional rotational accretion, which can facilitate a long-lived pair of energetic jets with a fixed axis, remains speculative. Its main aim is to use pre-collapse core rotation to produce a fixed-axis pair of jets inclined to the pre-collapse core angular momentum. The motivation comes from the morphology of SN 1987A. 
In particular, I note the following challenges it needs to address. 

(1) While in the early phases of the explosion, fluctuations dominate to avoid a bipolar explosion along the rotating axis of the binary system (along which the companion spun up the core), at late times the rotation set by the two opposite accreted zones should dominate (equation \ref{eq:jdir}). This might be possible if the accreted material that sets the pair of jets comes from a radiative layer in the pre-collapse core, where the seeds of angular momentum fluctuations are very small. 

(2) To prevent the fixed-axis explosion along the binary angular momentum (the triple-ring axis), the companion to SN 1987A could not have spun up the core to fast rotation (equation \ref{eq:ratios}). This implies that the accreted gas has too low specific angular momentum to form an accretion disk. It can form an accretion belt, which might also launch jets in CCSNe \citep{SchreierSoker2016} 

(3) The iron emission (e.g., \citealt{Larssonetal2016, Larssonetal2013, Larssonetal2023, Wessonetal2026}) follows the bipolar (Keyhole) molecular material in the south, but differs in the north (see Figure 3 in \citealt{Soker2026SN1987Amulecular}). If so, the jets that shaped the Keyhole played a role in the synthesis of nickel that decayed to iron. For the late directional rotational accretion to explain this iron distribution, the jets cannot be too late. They must be active while the core material (silicon and oxygen) is still near the center so that the jets can induce the nuclear reactions to form nickel.  

For these challenges, the late directional rotational accretion I proposed here is, as I indicated above, still a speculative scenario. A simple alternative explanation is that the SN 1987A's progenitor was born with a spin inclined to the binary angular momentum by $\simeq 45^\circ$, and that the inner core maintained this direction of core rotation until its collapse. This raises the question of why not all CCSNe show a clear bipolar structure.  

\section{Summary} 
\label{sec:Summary}

We are in the fortieth anniversary of SN 1987A. Shortly after its discovery, studies suggested that the SN 1987A progenitor experienced a common envelope evolution with a main-sequence companion to account for the asymmetrical explosion \citep{ChevalierSoker1989} and the blue progenitor \citep{Podsiadlowskietal1990A}. In this study, I addressed two aspects resulting from that binary interaction: the influence of the triple-ring bipolar system that the progenitor ejected $\simeq 20,000 \yr$ before the explosion on the light curve for a hypothetical equatorial observer, and the spin-up of the core by the companion at the end of the common envelope evolution. 

In Section \ref{subsec:OnlyRing}, I showed, under simple assumptions, that the equatorial ring has obscured the central star from a hypothetical observer in and near its plane. For hundreds of years, the bolometric luminosity would be $<1 \%$ of the central stellar luminosity if it were only from the equatorial ring (equation \ref{eq:Ltot}). However, as I showed in Section \ref{subsec:OuterRings}, the outer rings scattered $\simeq 5 \%$ of the stellar radiation towards this observer for a few hundred years. By that time, the equatorial ring became more transparent, and the luminosity slowly increased over a few thousand years (Section \ref{subsec:FailedEvent}). I argued there that this event would be classified by present terms as a ``failed supernova''. However, there was no core collapse, no black hole formation, and nothing failed. It was a type II ILOT (Section \ref{sec:intro}). The entire event might have started with a brightening, as the companion accreted mass, as some models suggest for many other ILOTs of different kinds (e.g., \citealt{Kashietal2020}), and launched jets that shaped the two outer rings, as supported by jet-shaped planetary nebulae (e.g., \citealt{BeltranSanchezetal2026}). Like this event, I argue that Type II ILOTs occur for massive stars. In some cases, the fading will be partial, as might have been the case with WOH G64, an RSG that suffered a fading (for recent studies of WOH G64, see, e.g., \citealt{MunozSanchezetal2026, vanLoonOhnaka2026}). 
The common occurrence of type II ILOTs might strengthen earlier claims that the ``failed supernova'' candidates N6946-BH1 and M31-2014-DS1 are type II ILOTs and not the collapse of a star to a black hole as in a ``failed supernova'' (Section \ref{sec:intro}).    

In Section \ref{sec:Misalignedaxes} I addressed the question of why the bipolar structure of SN 1987A, the Keyhole, as revealed, for example, by its molecular material (e.g., \citealt{Soker2026SN1987Amulecular}), is not aligned with the symmetry axis of the three circumstellar rings (Section \ref{subsec:Motivation}). There is the simple possibility that the SN 1987A progenitor had a natal spin inclined at $\simeq 45^\circ$ to the orbital angular momentum, and the core maintained it until collapse. I considered another scenario that assumes (Section \ref{subsec:Motivation}) that the companion spun up the core and that the pre-collapse core rotation played a role in forming the bipolar structure. I crudely estimated (Section \ref{subsec:SpunUpCore}) a plausible value for the spun-up pre-collapse core (equation \ref{eq:je}). I found that although the specific angular momentum is not sufficient to form an accretion disk fully supported by the centrifugal force, the rotation is non-negligible with respect to the required specific angular momentum to form a disk, and not with respect to the angular momentum fluctuations in the convective layers of the core (equation \ref{eq:ratios}).   
Based on this, I proposed the `late directional rotational accretion' scenario (Section \ref{subsec:LateTimes}). It has four main phases, as listed there. Mainly, the angular momentum fluctuations seeded by core convection dominated at the beginning of the explosion; several jittering jet pairs occurred during this phase. The explosion starts accelerating the core material in most, but not all, directions. Material that, at late times, accreted from mid-latitude directions with respect to the pre-collapse rotation might have sufficient angular momentum due to pre-collapse core rotation to form an accretion belt; i.e., the material has sub-Keplerian specific angular momentum. The angular momentum axis of this material, hence the accretion belt, is inclined to the pre-collapse angular momentum. The accretion belt might have launched the jets that shaped the Keyhole of SN 1987A. This scenario remains speculative, as it must overcome some challenges that I list at the end of Section \ref{subsec:LateTimes}.  

Earlier studies have used SN 1987A morphology to argue for the JJEM as its explosion mechanism and, by extension, to strengthen the JJEM as the primary explosion mechanism of CCSNe. In this study, I used the properties of the three rings to show that an observer in and near the equatorial ring would have seen SN 1987A progenitor luminosity decreasing for hundreds to thousands of years, as the ring ejection was a type II ILOT for such an observer. This shows that ``fail supernovae'' might be type II ILOTs, and the fraction of ``fail supernovae'' is much below the prediction of the neutrino-driven mechanism. I also used the JJEM to explain the misalignment between the three-ring axis and the SN 1987A bipolar structure, its Keyhole. Overall, this study adds small but unique support to the claim that the JJEM is the primary explosion mechanism of CCSNe.

\section*{Acknowledgments} 
A grant from the Pazy Foundation 2026 supported this research.
I thank the Charles Wolfson Academic Chair at the Technion. 





 \bibliography{reference}{}
  \bibliographystyle{aasjournal}
 
\end{document}